\documentclass[letterpaper, 10 pt, conference]{ieeeconf}  

\usepackage[utf8]{inputenc}
\usepackage{graphicx}
\usepackage{subfig}

\usepackage{amsmath, amsthm, amssymb}
\makeatletter
\let\NAT@parse\undefined
\makeatother
\usepackage[colorlinks,citecolor=blue,linkcolor=red]{hyperref}

\IEEEoverridecommandlockouts                              

\title{\LARGE \bf
Learning to Influence Vehicles' Routing in Mixed-Autonomy Networks by Dynamically Controlling the Headway of Autonomous Cars
}

\author{Xiaoyu Ma$^{1}$ and Negar Mehr$^{2}$
\thanks{*This work is supported by the National Science Foundation, under CAREER Award ECCS-2145134.}
\thanks{$^{1}$Xiaoyu Ma is with the Department of Electrical and Computer Engineering, UIUC, 306 N Wright St, Urbana, IL 61801, USA
        {\tt\small xiaoyum2@illinois.edu}}%
\thanks{$^{2}$Negar Mehr is with the Department of Aerospace Engineering, UIUC, 104 S Wright St, Urbana, IL 61801, USA
        {\tt\small negar@illinois.edu}}%
}

\begin{document}

\maketitle
\thispagestyle{empty}
\pagestyle{empty}

\begin{abstract}
It is known that autonomous cars can increase road capacities by maintaining a smaller headway through vehicle platooning. Recent works have shown that these capacity increases can influence vehicles' route choices in unexpected ways similar to the well-known Braess's paradox, such that the network congestion might increase. In this paper, we propose that in mixed-autonomy networks, i.e., networks where roads are shared between human-driven and autonomous cars, the headway of autonomous cars can be directly controlled to influence vehicles' routing and reduce congestion. We argue that the headway of autonomous cars --- and consequently the capacity of link segments --- is not just a fixed design choice; but rather, it can be leveraged as an {infrastructure control} strategy to {dynamically} regulate capacities. Imagine that similar to variable speed limits which regulate the maximum speed of vehicles on a road segment, a control policy regulates the headway of autonomous cars along each road segment. We seek to influence vehicles' route choices by directly controlling the headway of autonomous cars to prevent Braess-like unexpected outcomes and increase network efficiency. We model the dynamics of mixed-autonomy traffic networks while accounting for the vehicles' route choice dynamics. We train an RL policy that learns to regulate the headway of autonomous cars such that the total travel time in the network is minimized. We will show empirically that our trained policy can not only prevent Braess-like inefficiencies but also decrease total travel time\footnote{The code is available at: \\ \href{https://github.com/labicon/RL-Traffic-Dynamics}{\texttt{https://github.com/labicon/RL-Traffic-Dynamics}}}. 


\end{abstract}


\section{Introduction}

Road networks will soon be shared between human-driven and autonomous cars, i.e., they will operate under mixed vehicle autonomy. Unlike human-driven cars which cannot be externally controlled, we can assume a level of control over autonomous cars, which provides us with an additional control input to affect traffic networks. 
Several recent works have considered controlling autonomous cars' actions at a vehicle level for improving traffic conditions~\cite{wu2017emergent, darbha1999intelligent,yi2006macroscopic, mehr2018game,mehr2021game}. In this work, we focus on the system-level performance of mixed-autonomy networks. 
A particularly attractive feature of autonomous cars for influencing system-level mobility is that they can facilitate vehicle platooning. Vehicle platoons are groups of vehicles that can maintain a shorter headway using adaptive cruise control technologies. As a result, forming vehicle platoons can result in increased capacity of road segments~\cite{lioris2017platoons}. Such capacity increases can be up to threefold if all vehicles are autonomous~\cite{lioris2017platoons}.

While such capacity increases have the potential to increase traffic throughput and reduce congestion, it was shown in recent works that if all vehicles select their routes \emph{selfishly}, the capacity increases of autonomous cars might in fact \emph{worsen} congestion~\cite{mehr2018can,mehr2019will}. It is well known that traffic networks tend to operate at equilibria, where vehicular flows are routed along the
network paths such that no vehicle can gain any savings in its travel time by unilaterally changing its route~\cite{wardrop1952some}. Under such selfish route choices, in our previous
work~\cite{mehr2018can, mehr2019will}, we showed that the capacity increases of autonomous cars can influence vehicles' route choices such that the overall network delay may increase. This unexpected behavior is similar to the well-known Braess's paradox~\cite{braess1979existence} where it was shown that increasing a link capacity by adding lanes to the link can increase the overall network delay. 

Several recent works have focused on control schemes that alleviate the inefficiencies that result from selfish route choices of vehicles in mixed-autonomy networks. For example,~\cite{mehr2019pricing,lazar2019optimal,biyik2021incentivizing} have considered tolling mixed-autonomy traffic networks such that network inefficiencies are minimized. In~\cite{biyik2018altruistic,lazar2021learning,kolarich2022stackelberg}, the altruistic routing of autonomous cars was studied for controlling mixed-autonomy networks, where the routing of autonomous cars was controlled such that when human-driven cars react selfishly to the routing of autonomous cars, the overall network performance is improved. In all these works, the assumption is that autonomous cars maintain platoons of cars with a pre-specified headway, and control mechanisms are sought to alleviate the inefficiencies that arise from selfish routing.

\begin{figure}[t]
    \centering
    \includegraphics[width=0.5\textwidth]{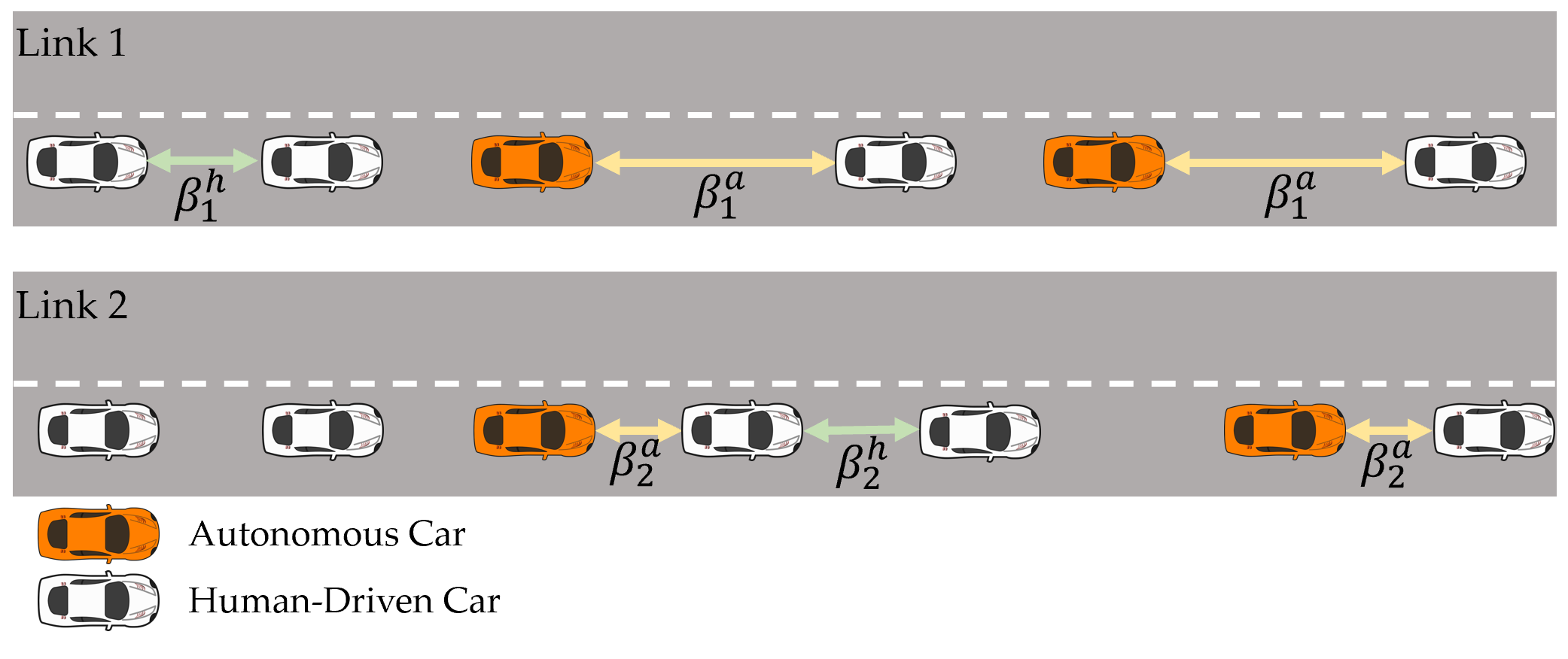}

    \caption{We propose that we can dynamically control the headway of autonomous cars $\beta^a_1$ and $\beta^a_2$ on road segments to influence the vehicles' routing and reduce congestion. We assume that the headway of human-driven cars ($\beta^h_1$ and $\beta^h_2$) remains uncontrolled and constant.}%
    \label{pic_dgm}
    \vspace{-7mm}
\end{figure}

In this paper, we take a different approach and propose that in mixed-autonomy networks, the capacity increases of autonomous cars can be \emph{directly regulated} to reduce network inefficiencies and improve performance. We argue that the headway of autonomous cars--- and consequently the capacity increases of autonomous cars--- is not just a fixed design choice; but rather, it can be leveraged as an \emph{infrastructure control} strategy to \emph{dynamically} regulate road capacities (see Fig.~\ref{pic_dgm}). Similar to variable speed limits, which regulate the maximum speed of vehicles on a road segment, a control policy can regulate the headway of autonomous cars along each road segment. This is in fact closely related to variable speed limits which have been widely studied for networks of only human-driven cars, where link capacities are regulated by dynamically regulating the maximum speed along network links~\cite{lin2004exploring,khondaker2015variable,muralidharan2012optimal,bertini2006dynamics}. In this work, we seek to influence vehicles' route choices by directly controlling the headway of autonomous cars to prevent Braess-like unexpected outcomes and increase network efficiency. 


To investigate the impact of our proposed controller, we model the dynamics of mixed-autonomy traffic networks while accounting for the vehicles' route choice dynamics. Since traffic dynamics and vehicles' choice dynamics are highly nonlinear, finding an appropriate controller is extremely challenging. To tackle this, we propose to train an RL policy that learns to regulate the headway of autonomous cars such that the total travel time in the network is minimized. We will show through empirical analysis that not only can our trained policy prevent Braess-like inefficiencies but also it can decrease total travel time and improve network performance. Our evaluations reveal that dynamically regulating the headway of autonomous cars has the potential to reduce the inefficiencies of selfish routing in mixed-autonomy traffic networks. In summary, our main contributions are:

\begin{itemize}
	\item We propose to dynamically regulate the headway of autonomous cars to influence vehicles' route choices. 
	\item We use traffic dynamics and vehicles' route choice dynamics to train an RL policy for dynamically controlling the headway of autonomous cars.
	\item We demonstrate empirically through our experiments that such an RL policy can reduce the inefficiencies that are inherent in selfish route choices and can improve the performance of mixed-autonomy traffic networks. 
\end{itemize}



\section{Network Model}\label{sec:model}

We model the traffic network as a directed graph $\network = (\nodeset, \linkset)$ where $\nodeset$ is the set of nodes and $\linkset$ is the set of links. 
Each link $\link \in \linkset$ in the network is a road segment which is a directed edge from a node $u \in \nodeset$ to a node $v \in \nodeset$. We assume that we are given a set of origin-destination (O/D) pairs denoted by $\ODset$, which contains $k$ O/D pairs. Each O/D pair $w \in \ODset$ is denoted as $(o_\OD, d_\OD)$,  where $o_\OD, d_\OD \in \nodeset$ ($o_\OD \neq d_\OD$) represent the origin and destination nodes of O/D pair $w$ respectively. We define the vector of network demand to be $\demand(t) = (\demand_\OD(t))_{\OD \in \ODset}$ where $\demand_\OD(t)$ is the potentially time-varying demand profile of OD pair $\OD$. Note that $\demand_\OD(t)$ specifies the total amount of vehicular flow (including both human-driven and autonomous vehicles) that needs to be routed along O/D pair $w$ at time $t$.

For each O/D pair $\OD \in \ODset $, we assume that there exists at least one path from its origin to its destination. We use $\pathset_\OD$ to denote the set of all paths that connect O/D pair $\OD$. Additionally, let $\pathset = \cup_{\OD \in \ODset} \pathset_w$ denote the set of all network paths. 
For each $\path \in \pathset$, we denote the time-dependent flow of autonomous and human-driven cars along $p$ by $\auto_\path(t)$ and $\human_\path(t)$, respectively. We further define the vectors of autonomous and human-driven flows as $f(t) := (f^h_p(t), f^a_p(t))_{p \in \mathcal{P}}$, where $\auto(t) = (\auto_\path(t))_{\path \in \pathset}$ and $\human(t) = (\human_\path(t))_{\path \in \pathset}$. We define the total flow on a link $\link \in \linkset$ at time $t$ as $f_\link(t) = f_\link^a(t)+f_\link^h(t)$. We also define the autonomy fraction along a link $\link \in \linkset$ to be the fraction of cars along the link that are autonomous, i.e. $\alpha_\link(t) := \frac{f_\link^a(t)}{f_\link^a(t) + f_\link^h(t)}$.

\begin{figure}
      \centering
      \includegraphics[width=0.4\textwidth]{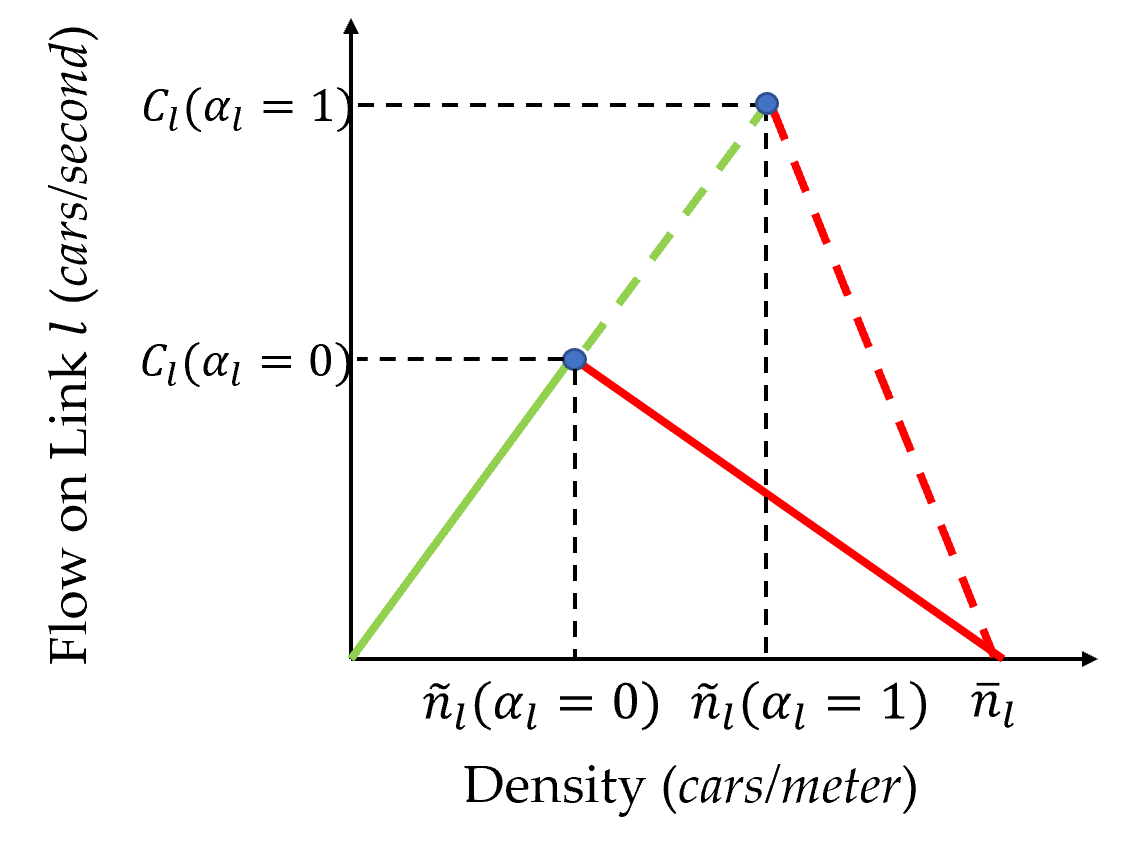}
      \caption{The fundamental diagram of mixed-autonomous traffic for varying autonomy fractions $\alpha_\link$ along link $\link$. As the fraction of autonomous vehicles $\alpha_l(t)$ increases, the link capacity $C_l(\alpha_l)$ also increases while the jam density $bar{n}_l$ is constant. }
      \label{fig:fundamental}
      \vspace{-6mm}
\end{figure}

\subsection{Capacity Model}
We assume that the flow along each link is governed by the fundamental diagram of the link~\cite{cassidy2011macroscopic}, i.e. the maximum flow along a link is the link's capacity, and the maximum density of the vehicles along a link is the jam density. Unlike networks with only human-driven cars where the link capacity is constant, when autonomous cars with variable headways are deployed, a link capacity changes as a function of both the headway of autonomous cars and the autonomy fraction along the link.
Let the headway distance maintained by autonomous cars and human-driven cars on link $\link \in \linkset$ at time $t$ be $\beta_\link^a (t)$ and $\beta_\link^h (t)$ respectively. In the following, we discuss how the capacity changes of a link can be modeled.

We let $\critdens_\link(\alpha_\link(t))$ denote the critical density of link $\link$ at time $t$. Critical density determines when the traffic on the link changes from free flow to congestion. Similar to link capacity, the critical density $\critdens_\link(\alpha_\link(t))$ is a function of the autonomy fraction along the link $\alpha_\link$. The critical density is the maximum density of vehicles that can travel on link $l$ in free flow (see Fig.~\ref{fig:fundamental}). It was shown in~\cite{lazar2021learning,biyik2021incentivizing} that in mixed-autonomy networks, the critical density of a link is equal to
\vspace{-3mm}
\begin{align}\label{eq:critical-density}
    \critdens_\link(\alpha_\link(t)) =  \frac{b_\link}{\alpha_\link(t) \cdot \beta_\link^a(t) + (1-\alpha_\link(t)) \cdot \beta_\link^h(t)},
\end{align}
where $b_\link$ is the number of lanes on link $\link$.
The weighted sum of headways in the denominator of Equation~\eqref{eq:critical-density} represents the average headway kept between one car and its preceding car when all vehicles are running in free flow. Let the capacity of a link $\link$ at time $t$ be denoted by $C_\link(\alpha_\link(t))$. It can be then shown that the capacity of link $\link$ is $C_\link(\alpha_\link(t)) = \freeflowvel_\link \cdot \critdens_\link(\alpha_\link(t))$, where $\freeflowvel_\link$ is the free-flow velocity on link $\link$. The jam density, denoted as $\jamdens$, is the density on a link when the link gets totally jammed and the speed of vehicles running in the link reduces to zero, and is assumed to be a constant. 

The density of a link $l$ at time $t$ is defined as $n_\link(t)=n_\link^a(t)+n_\link^h(t)$, where $n_\link^a(t)$ and $n_\link^h(t)$ are the density of autonomous and human-driven cars respectively. Following flow conservation, for any link $\link \in \linkset$, at every time step $t$, we have: 
\vspace{-3mm}
\begin{align}\label{eq:link_update}
    n_\link(t+1) = n_\link(t) + f_{\link,in}(t) - f_{\link,out}(t).
\end{align}
where $f_{\link,in}(t)$ and $f_{\link,in}(t)$ are the input and output flows of link $\link$ at time $t$.

Following the fundametnal diagram, the flow of each link $\link \in \linkset$ at time $t$ can be calculated as
\begin{align}\label{eq:flow}
    f_\link(n_\link(t)) = \begin{cases}
            \freeflowvel_\link \cdot n_\link(t), &\text{if }n_\link(t) \leq \critdens_\link(\alpha_\link(t)) \\
            \frac{\freeflowvel_\link \cdot \critdens_\link(\degauto_\link(t)) \cdot (\jamdens_\link - n_\link(t))}{\jamdens_\link - \critdens_\link (\alpha_\link(t))}, &\text{if } \critdens_\link (\alpha_\link(t)) \leq n_\link(t) \leq \jamdens_\link \\
            0, &\text{else}
        \end{cases}.
\end{align}
Equation~\eqref{eq:flow} indicates that when the density $n_\link(t)$ is smaller than the critical density, vehicles move at the free-flow speed. When $n_\link(t)$ is larger than the critical density, the vehicles are in the congested regime as shown in Fig.~\ref{fig:fundamental}.

\subsection{Latency Function}
For each link $\link$ in $\linkset$, we define a binary parameter $s_\link(t)$ indicating whether the link is in free flow at time step $t$ ($s_\link(t)=0$) or congested ($s_\link(t)=1$). Then, using the results from~\cite{biyik2021incentivizing}, the latency along a link $l \in \linkset$ can be found by:
\begin{align}\label{eq:latency}
    e_\link(f_\link(t), s_\link) = \begin{cases}
            \frac{d_\link}{\freeflowvel_\link}, &\text{if }s_\link(t)=0 \\d_\link(\frac{\jamdens_\link}{f_\link(t)}+\frac{\critdens_\link(\alpha_\link(t))-\jamdens_\link}{\freeflowvel_\link \critdens_\link(\alpha_\link(t))}), &\text{if }s_\link(t)=1
            \end{cases},
\end{align}        
where $d_\link$ is the length of link $\link$. Eq.~\eqref{eq:latency} captures the time that it takes for one vehicle to travel along the link. Consequently, for each path $p \in \pathset$, the path latency $\latency_\path$ is the sum of link latency along the links that constitute path $p$:
\begin{align}\label{eq:latency_p}
    \latency_\path = \sum_{\link' \in \path} \latency_\link' (f_{l'}(t))
\end{align}   

\subsection{Route Choice Dynamics}
To capture the impact of the headway of autonomous cars on the vehicles' route choices, we need to model the choice dynamics of vehicles in the network. We model this through evolutionary dynamics introduced in~\cite{lazar2021learning} to model how a population of vehicles along a path $p \in \pathset$ choose their route. More specifically, for a path $p$ connecting O/D pair $\OD$, we have:
\begin{align}\label{eq:greedy_human}
    \human_\path(t+1) = 
    \begin{aligned}
        \frac{\human_\path(t)\exp(-\mu^h \latency_\path(t))}{\sum_{\path' \in \pathset_\OD} \human_{\path'}(t) \exp(-\mu^h \latency_{\path'}(t))},
    \end{aligned}
\end{align} 
\begin{align}\label{eq:greedy_auto}
    \auto_\path(t+1) = \begin{aligned}\frac{\auto_\path(t)\exp(-\mu^a \latency_\path(t))}{\sum_{\path' \in \pathset_\OD} \auto_{\path'}(t) \exp(-\mu^a \latency_{\path'}(t))},
    \end{aligned}
\end{align} 
where $\mu^h$ and $\mu^a$ are the rationality factors for human-driven and autonomous vehicles that capture the selfishness of vehicles when selecting their route.  
In~\eqref{eq:greedy_human} and~\eqref{eq:greedy_auto}, the fraction of vehicles running on path $\path$ at time step $t+1$ is considered to be inversely proportional to the exponential of the delay experienced by users of that road. The parameters $\mu^h$ and $\mu^a$ were introduced in previous work as learning rates in the context of humans’ routing choices and simulate a congestion game~\cite{lam2016learning}. Intuitively, when the value of $\mu^h$ or $\mu^a$ is large, it is more likely for the group of corresponding vehicles to actively change their route due to a change in the latency of their previously selected route.


\section{Learning to Adjust the Headway of Autonomous Cars}

Equipped with a model that can simulate the impact of autonomous cars' headway on the route choice of vehicles, we can now train a policy that learns to control the headway of autonomous cars such that when all cars select their routes greedily, the overall performance of the network is improved. We discuss how we train a policy to achieve this goal.

\subsection{Reward function}
An intuitive reward function for training the policy is the negative of the Total Travel Time (TTT) of the network. The network TTT is defined as the sum of vehicle densities across all network links over a finite horizon of time $T$.  
    \vspace{-2mm}
    \begin{align}\label{eq:TTT}
       r(T) = -\sum_{t=1}^T\sum_{\link \in \linkset} n_\link(t).
    \end{align}
    Since we are trying to alleviate congestion, maximizing~\eqref{eq:TTT} can be viewed as minimizing the total densities of all vehicles in the network. This choice is motivated by the common use of TTT in the transportation literature to capture network performance (see~\cite{gomes2006optimal} for a list of references).

\subsection{Policy Training}
We would like to find a policy $\pi$ that outputs the headway of autonomous cars along each link $\beta_l^a$ such that the reward~\eqref{eq:TTT}
is maximized when vehicles choose their routes according to~\eqref{eq:greedy_human} and~\eqref{eq:greedy_auto}. We assume that we cannot control the headway of human-driven cars, and we have control only over the headway of autonomous cars along each link $\beta_l^a$. However, this is an extremely challenging control problem since the traffic and route choice dynamics of the vehicles are highly nonlinear and complex. To tackle this challenge, we propose to train a policy that learns to adjust the autonomous cars' headway for minimizing TTT. We train our policy in a model-free setup where we leverage our traffic dynamics model to develop a traffic simulator capable of capturing the dependence of vehicles' route choices on vehicles' headway. We used this as a simulation platform for training our RL agent that learns to adjust the headway of autonomous cars over a fixed time horizon $T$. 


The problem can be incorporated as a finite-horizon discounted Markov decision process (MDP), defined by the tuple $(\mathcal{S}, \mathcal{A}, P, c, \rho_0, \gamma)$,
where $\mathcal{S}$ is the set of states, $\mathcal{A}$ is the set of actions,
$P : \mathcal{S}\times\mathcal{A}\times\mathcal{S}\rightarrow\mathbb{R}$ is the transition probability distribution, $r : \mathcal{S}\rightarrow\mathbb{R}$ is the reward function, $\rho_0 : \mathcal{S} \rightarrow \mathbb{R}$ is the distribution of the initial state $s_0$ which is equal to the initial density of the vehicles, and  $\gamma\in (0, 1)$ is the
discount factor. The observation space is the same as the state space $S \in \mathcal{S}$, which includes the density of vehicles on the road segments, while the action space $A \in \mathcal{A}$ is the headway of autonomous cars along the links at each time step $(\beta_\link^\alpha(t))_{\link \in \linkset}$. The transition function is determined by the system dynamics~\eqref{eq:link_update},~\eqref{eq:greedy_human}, and~\eqref{eq:greedy_auto}.

Our implemented simulator can be used in conjunction with state-of-the-art RL algorithms. We selected Proximal Policy Optimization (PPO)~\cite{schulman2017proximal} to train our RL agent. Since our state and action spaces are continuous, we chose PPO which is a policy-gradient-based method. PPO dynamically adjusts the policy $\pi$ according to the gradient of the reward with respect to the policy parameters. The clipping mechanism in PPO can protect the policy from over-updating itself with respect to large and steep gradients.

\begin{figure}
    \centering

    \subfloat[]{{\includegraphics[width=0.25\textwidth]{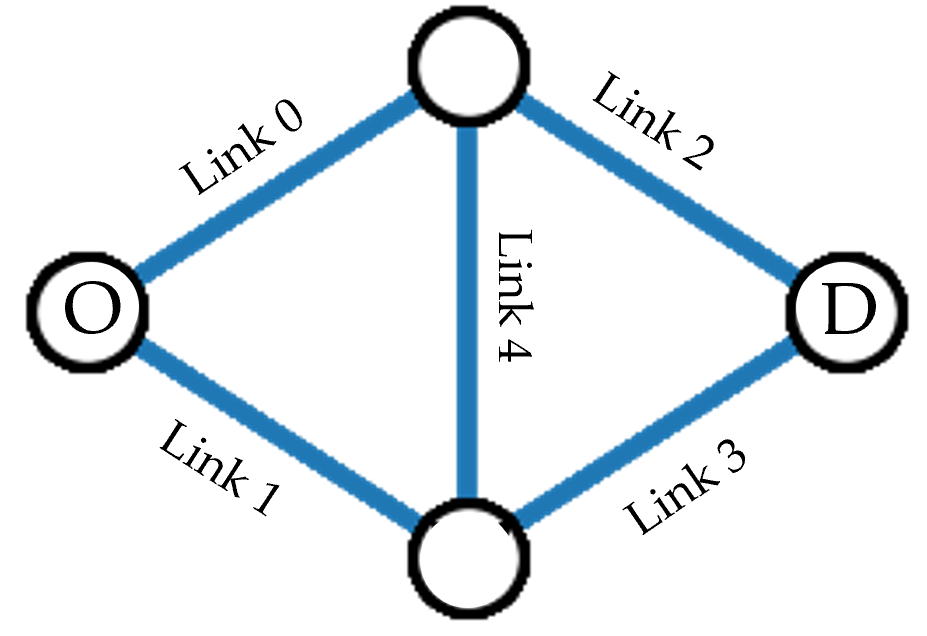}}}
    \subfloat[]{{\includegraphics[width=0.25\textwidth]{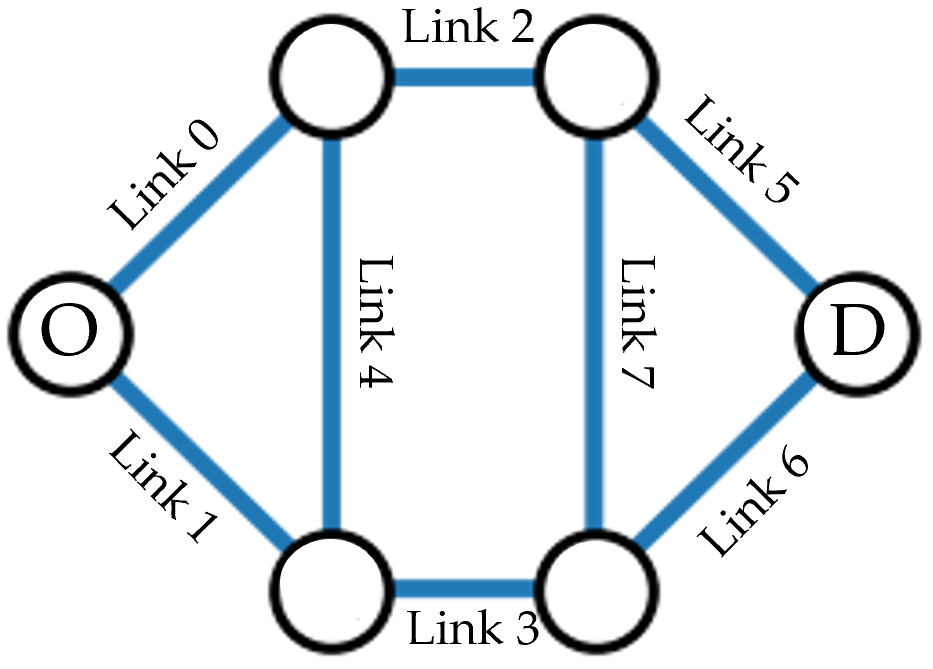}}} 

    \caption{The geometry of the classic Braess network (a) and the extended 8-link network with embedded Braess geometry (b). }
    \label{fig:pic_braess}
    \vspace{-8mm}
\end{figure}

\section{Experiment}

\subsection{Experiment Design}

In this section, we introduce the details of our experimental setup. We will start with a Braess network geometry shown in Fig.~\ref{fig:pic_braess} (a). We will then consider a larger network with 8 links that has the Braess geometry embedded in it as shown in Fig.~\ref{fig:pic_braess} (b). Note that the Braess geometry is a very complicated network geometry that was shown in our previous work that can result in counter-intuitive vehicles' route choices~\cite{mehr2019will}. For example, in~\cite{mehr2018can}, it was shown that in a Braess network geometry, the capacity increases that result from vehicle platooning on the middle link (Link 4) can indeed increase congestion levels when vehicles select their routes selfishly. We chose the Braess network to verify if our control policy can prevent such unexpected behaviors and even more, improve the performance. 

We considered a network where Link 0, 1, 2, and 3 are designed to be 240 kilometers long while the middle link is 60 kilometers long. The links are designed to have a free-flow velocity of 30 meters per second. Each time step lasts for one minute. We assume that the headway of autonomous cars can be changed every 10 minutes, and the total length of the time sequence is 200 minutes. We chose each time step to last for 10 minutes accordingly. Although higher frequency headway changes are feasible, we wanted to see if a significant improvement in traffic performance can be sought even with less frequent updates on headways. We set the number of lanes on Link 0 and 3 to be two times larger than those on Link 1 and 2. The number of lanes on Link 4 is designed to be especially large so that we can better capture the impact of capacity increases. For the initial conditions, we put more vehicles onto Links 0 and 2 at the beginning of the time horizon so that it is more crowded on the top route of the network. For the 8-link Braess network, we keep Links 0 and 4 to be the same as the classic Braess network, while Link 5 and 7 are designed to have the same characteristics as Links 2 and 4.

\begin{figure}
    \centering
    \includegraphics[width=0.45\textwidth]{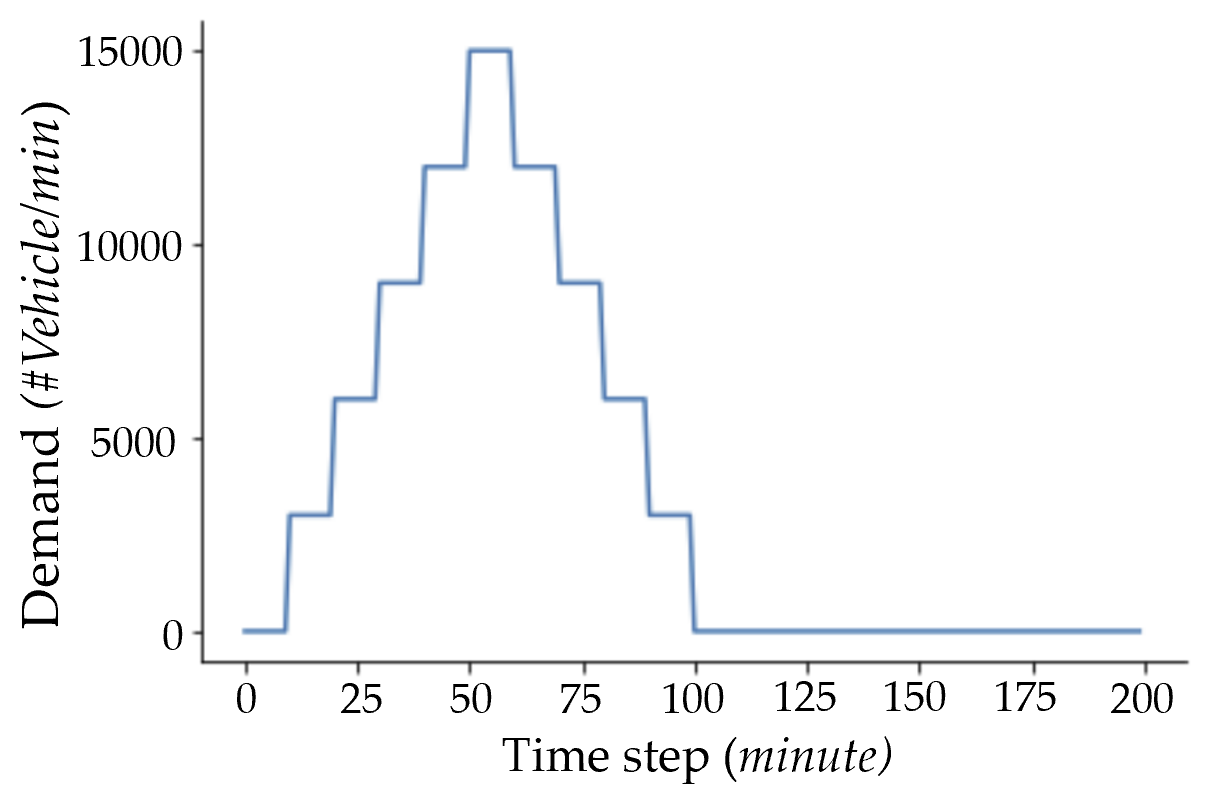}

    \caption{The time-dependent demand injected at node $O$ for both the classic Braess network and the 8-link extended Braess network.}
    \label{fig:pic_demand}
    \vspace{-6mm}
\end{figure}

We considered a time-varying vehicular demand entering the network through node $O$. We chose the demand profile such that it resembles the peak-hour demand profiles where the demand increases to a peak demand rapidly and then decreases (see Fig.~\ref{fig:pic_demand}). Such a choice allows us to verify our controller under realistic demand profiles which may vary significantly over time. We also consider a cool-down period in our demand profile similar to~\cite{gomes2006optimal} to evaluate the capability of our control policy in discharging vehicles that remained on the network. The simulator will reset itself to initial values and start a new training sequence after 20 consecutive episodes. The autonomous cars' headway is constrained to remain between 1 meter to 10 meters.

\begin{figure*}[h]
    \subfloat[]{\includegraphics[width=0.48\textwidth]{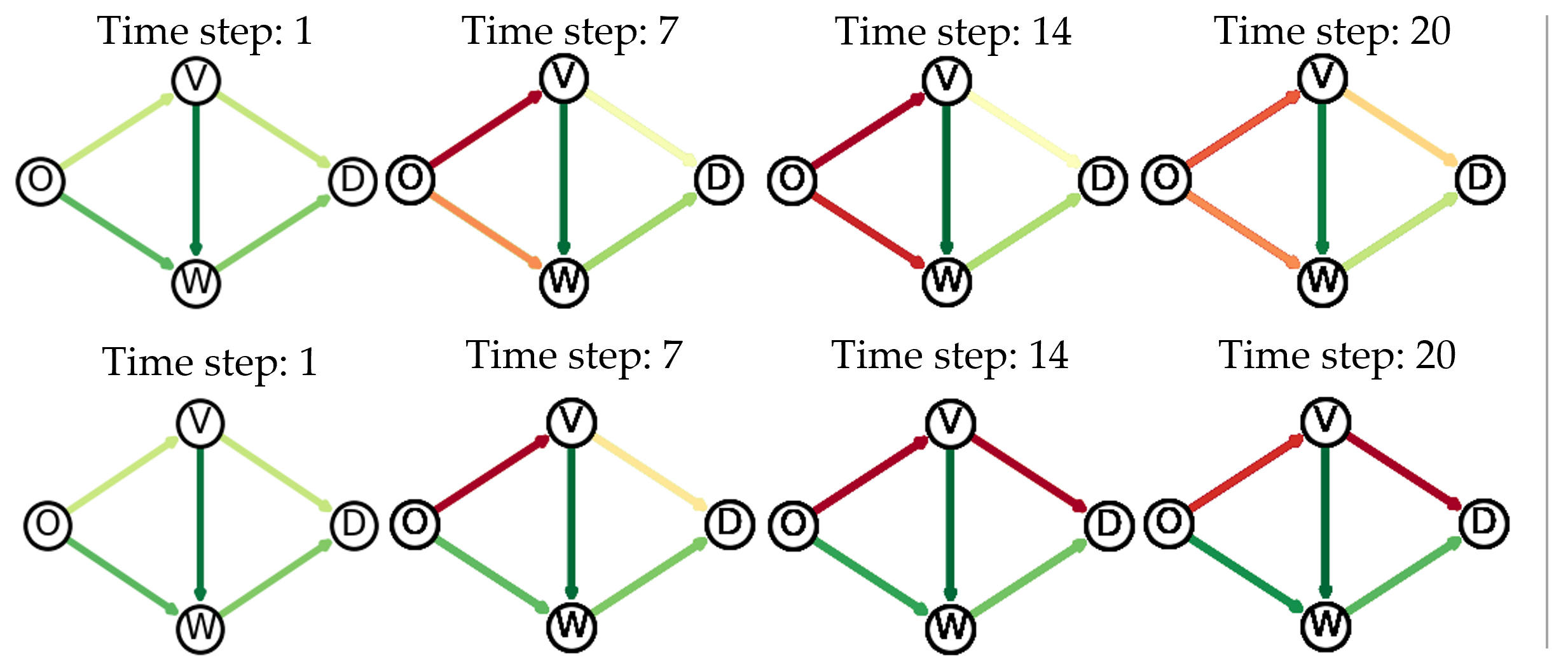}}
    \subfloat[]{\includegraphics[width=0.5\textwidth]{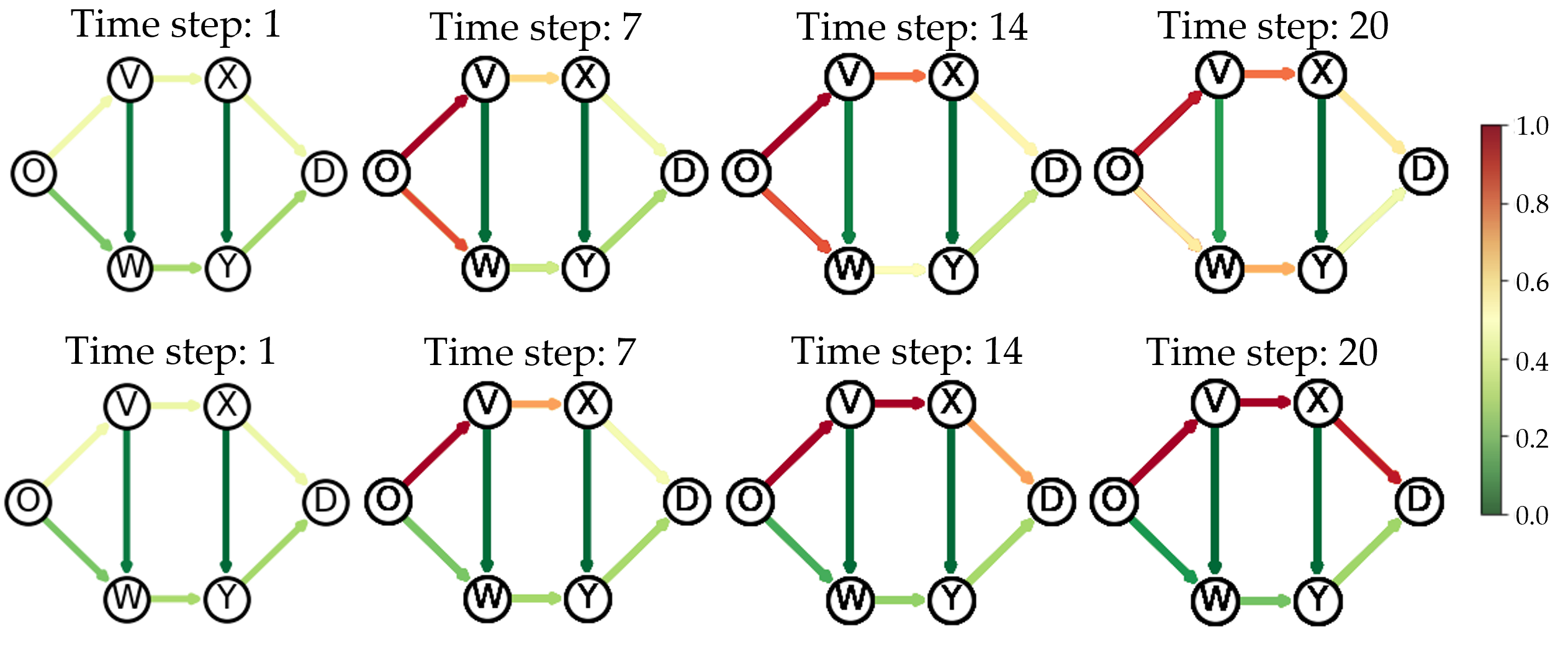}}

    \caption{Heat maps of vehicle densities that result from two policies on the Braess network (a) and extended 8-link Braess network (b). The top rows are the heat maps of the network in which autonomous cars use the same constant headway coefficient as human-driven cars at all times. In contrast, the bottom rows are heat maps of the networks in which the RL agent regulates the headway of autonomous cars. By changing the headway of autonomous cars dynamically, at least one route in the network remains in free flow, i.e. congestion is not propagated throughout the entire network. }%
    \label{fig:heatmap}
    \vspace{-7mm}
\end{figure*}

\subsection{Training Details}
Our experiments were carried out on Google Colab. The simulated environment is scripted in a Python environment connected with the baseline RL agent available in the StableBaselines library \cite{stable_baselines}. We chose a learning rate of $2\times 10^{-4}$, and set the number of steps to run for each environment per update to be 2048, the batch size to be 64, and the clipping range to be 0.2. Training the RL agent on the 5-link classic Braess Network typically takes at least 4 million time steps to reach convergence, and it will take approximately 5 hours of training on CPU. The 8-link extended Braess Network takes around 10 million time steps to converge, and it will take approximately 9 hours to finish. Note that the training time depends on the hardware available. The CPU that we used while training our policy on Google Colab was AMD EPYC 7B12.


\section{Results and Analysis}

In this section, we report the  performance of our trained policy. We compare the network TTT under our trained policy as opposed to two baselines which treat the headway of autonomous cars as constant. 1) Uniform constant headway: the baseline where the headways of both human-driven and autonomous cars are constant and equal. 2) Minimum Headway: the baseline where the headway of autonomous cars is set to be the minimum possible feasible headway. The Minimum Headway baseline was selected to investigate how the minimum headway and consequently maximum capacity increase of autonomous cars will affect vehicles' route choices. In comparison against these baselines, we also discuss some of the important factors that can affect the system's performance including the fraction of autonomous cars in the demand and the vehicles' rationality level captured in vehicles' rationality factors $\mu^a$ and $\mu^h$.

\subsection{Performance Comparison}

We apply our policy to the 5-link network for a horizon of $200$ minutes. We measure TTT and compare it with TTT over the same time horizon under our baseline methods. Table~\ref{table_perform} shows the percentage improvements in performance in terms of TTT when $\mu=0.1$ and the fraction of autonomous cars in the overall demand is equal to $\alpha_{O-D}=0.8$. As Table \ref{table_perform} demonstrates, not only can our policy avoid the unexpected increases in TTT due to the capacity increases of autonomous cars when using minimum headway, but also it improves the performance compared to the case when no vehicles' headway is controlled and are all equal to the headway of human-driven cars. Using the headway coefficients produced by our trained policy, the TTT is $4.0\%$ better than the network using uniform constant headway and is $11.1\%$ better than the network using minimum headway. This indicates that as demonstrated in our prior work~\cite{mehr2018can}, simply requiring autonomous cars to maintain platoons of vehicles is not necessarily going to improve network performance, it may actually worsen the performance when vehicles select their routes greedily as too many vehicles may be incentivized to change their routes in response to the capacity increases of autonomous cars. We ran a similar experiment with the 8-link network and found that our trained policy resulted in $2.8\%$ of improvement in TTT compared to the policy using uniform constant headway, and $0.2\%$ better than the policy using minimum headway. The percentage improvement is smaller in the 8-link network mainly due to the insufficient discharge of vehicles, as the time required for vehicles to exit the network is inherently longer than the classic Braess network.

\begin{table}[h]
\centering
\vspace{-3mm}
\begin{tabular}{|c||c|c|}
\hline
 &  Improvement w.r.t. & Improvement w.r.t.\\
 & Baseline 1	& Baseline 2 \\
\hline
Braess Network& 4.03\% & 11.16\% \\
\hline
8-link Network& 2.80\% & 0.20\% \\
\hline
\end{tabular}
\caption{The performance improvement of our policy compared to two constant headway policies: 1) uniform constant headway and 2) constant minimum headway.} 
\label{table_perform}
\vspace{-5mm}
\end{table}

To further show the difference in the system trajectories, Fig.~\ref{fig:heatmap} provides a comparison of the vehicles' distribution in the form of a heat map where a relatively congested link is painted in red, while a link in free flow is painted in green. We can see from the top row of these figures that with no headway control, both Link 0 and Link 1, end up congested. However, by changing the headway of autonomous cars dynamically (bottom row), at least one route in the network remains in free flow, i.e., the congestion is not propagated throughout the entire network. This allows the vehicles to have at least one route available to exit the network, thus improving the performance in terms of TTT.  


In our experiments, we found that two critical factors play essential roles in the effectiveness of the trained policy. One is the vehicles' rationality parameter $\mu$, representing the level of vehicles' selfishness in their route choice dynamics. The other is the total fraction of autonomous cars in the network. In the following, we will discuss the impact of these parameters.




\subsection{Vehicles' Rationality}
The rationality parameters $\mu^a$ and $\mu^h$ are coefficients that appear in Eq.~\eqref{eq:greedy_human} and~\eqref{eq:greedy_auto}, and they affect the portion of vehicles that change their route due to a change in the latency of their selected routes. Generally speaking, the larger $\mu^h$ and $\mu^a$ are, the more the vehicles will be likely to change their path to take a path with lower latency. Consequently, $\mu^a$ and $\mu^h$ will have a significant impact on how much one can influence vehicle' routing by controlling the headway of autonomous cars. 

\begin{figure}[t]
    \centering
    \includegraphics[width=0.45\textwidth]{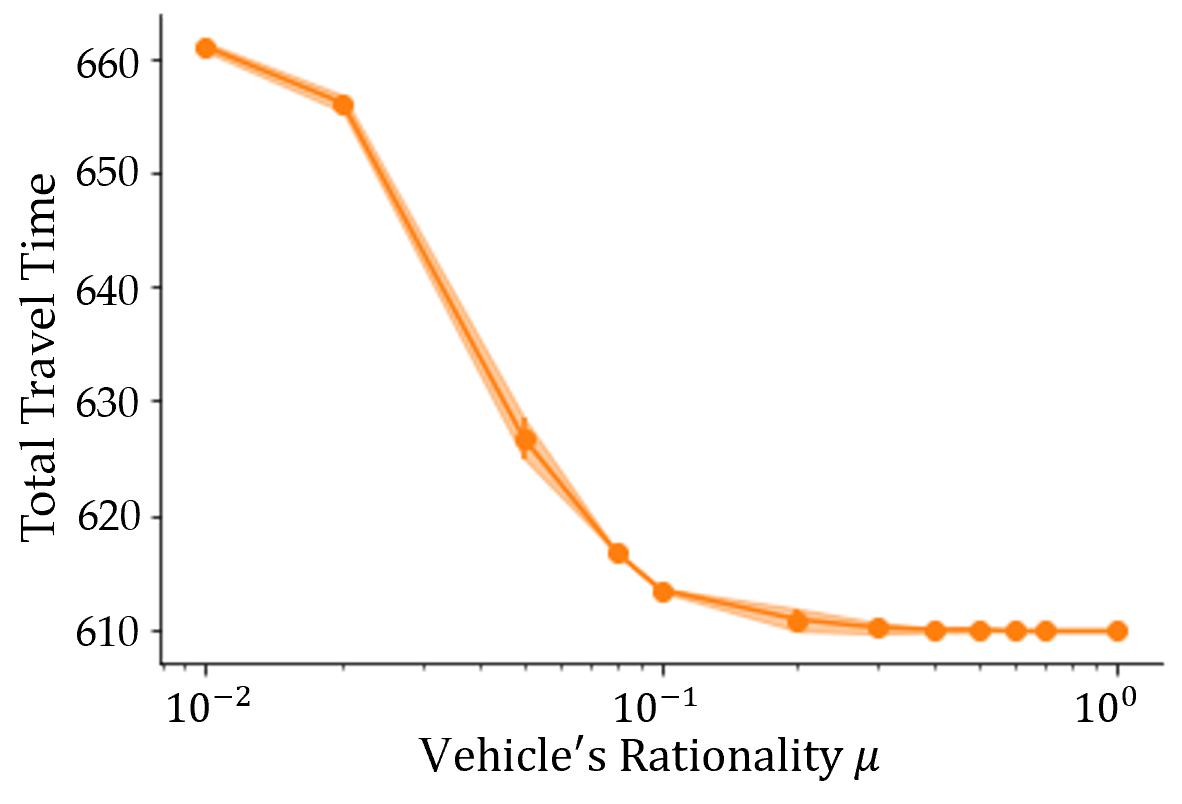}

    \caption{Plots of TTT as a function of the vehicles' rationality. We assume that the rationality coefficient of human-driven and autonomous cars are equal, i.e. $\mu^a=\mu^h=\mu$. The improvement in TTT is more significant with larger $\mu$. Note that the x-axis is plotted in a log scale.}%
    \label{pic_mu}
    \vspace{-6mm}
\end{figure}

To study this impact, we measured TTT under our trained policy for different levels of vehicles' rationality. We tried 5 random seeds for each rationality parameter, and the experiment is carried out on the classic Braess network. The resulting TTT for various rationality parameters is shown in Fig. \ref{pic_mu}. As the figure demonstrates, as the vehicles' rationality increases, TTT decreases, i.e. the impact of autonomous vehicles' headway on the route choices of vehicles become more significant.
Interestingly, the profile of links' headway generated by the RL agent differs for different levels of rationality. For example, the RL agent is more likely to generate smaller headways when using smaller $\mu^a$ and $\mu^h$, i.e, less selfish drivers, so that the critical capacity of the links can be increased more dramatically, which will help the policy provide larger incentives for vehicles to change their routes. When dealing with larger $\mu^a$ and $\mu^h$, the RL agent will assign larger headways to some links so that these links seem more crowded to the vehicles away from those links. In other words, the less selfish and greedy the vehicles are, the more the RL policy leverages the capacity increases of autonomous cars. This indicates that we need to carefully infer and estimate vehicles' selfishness when deriving a control strategy, as it can cause a significant difference in the resulting policies.

\begin{figure}[t]
    \centering
    \includegraphics[width=0.48\textwidth]{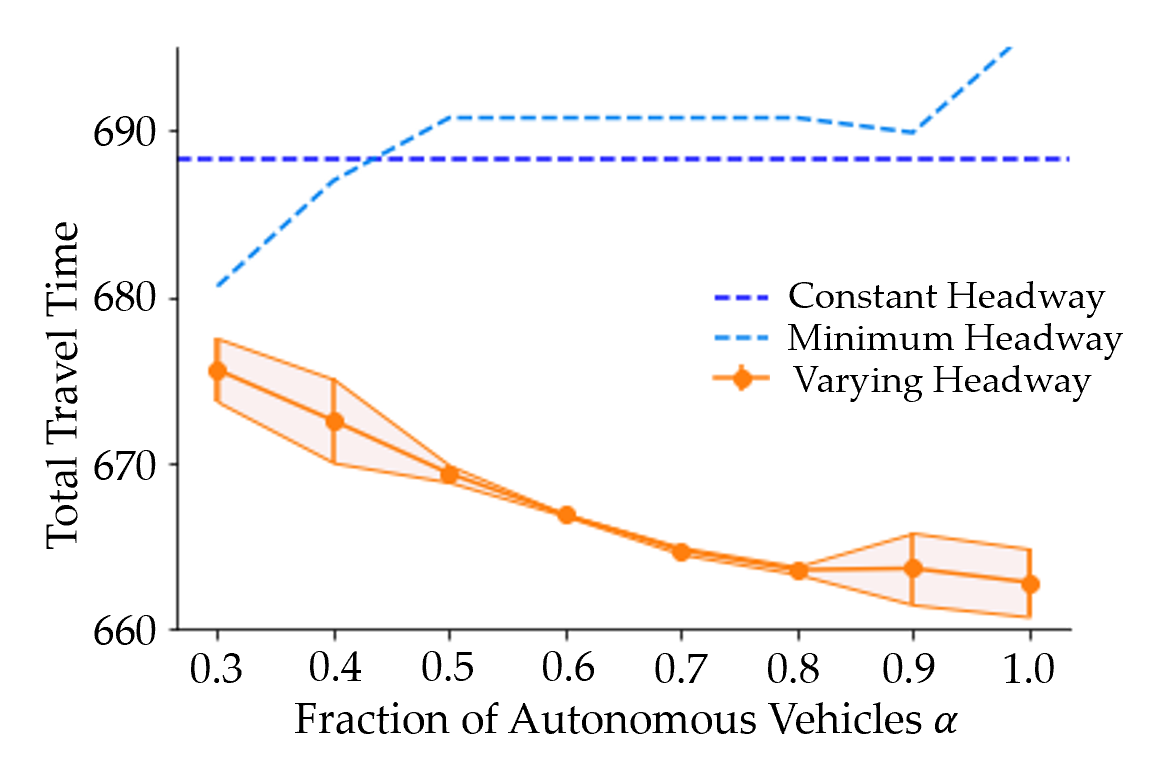}
    \caption{Plot of TTT as a function of the demand autonomy fraction $\alpha_{O-D}$, acquired by using the RL agent trained for each corresponding $\alpha_{O-D}$ separately. As the plot shows, the higher the fraction of autonomy is, the larger the savings in TTT are. Baselines using uniform constant headway and minimum headway are shown in dashed lines.}%
    \label{pic_alpha}
    \vspace{-7mm}
\end{figure}

\subsection{The total fraction of autonomous cars}
The total fraction of autonomous vehicles in the network will also inarguably affect the performance of our policy. The larger the fraction of autonomy is, the more control authority we will have for influencing vehicles' route choices. To study this, we keep the total number of vehicles constant and only vary the fraction of vehicles that are autonomous along O/D pair O-D denoted by $\alpha_{O-D}$. We varied the value of $\alpha_{O-D}$ and trained our policy to learn a policy tailored to the particular choice of autonomy fraction $\alpha_{O-D}$. Improvements in TTT against baselines using uniform constant or minimum headway are plotted in Fig.~\ref{pic_alpha}.

To examine the policies learned for each $\alpha_{O-D}$, we trained our policy on each $\alpha_{O-D}$ at least 5 times to compute TTT.  Overall, the performance of our policy keeps elevating when increasing the fraction of autonomous vehicles in the total demand. This is intuitive since effectively, the portion of the flow that we can control keeps growing. We see that for lower values of $\alpha_{O-D}$, the variance in TTT is larger. This is because, in the low $\alpha_{O-D}$ regime, the change in the headway maintained by autonomous vehicles is not having a significant effect on the traffic dynamics.

\section{Conclusion and Future Work}

In this work, we examined the feasibility of relieving traffic congestion by dynamically controlling the headway of autonomous vehicles. We modeled traffic networks with time-varying route choice dynamics.
We used this model to train a policy for dynamically controlling the headway of autonomous cars on the network links to improve the efficiency of the network. We measured the performance of our policy and demonstrated that our trained policy reduces total travel time. Our results indicate that controlling the headway of autonomous cars can be utilized to influence vehicles' route choices and reduce overall delays. 
In this work, we experimented on a relatively small-scale network due to an underpowered training platform. For our future work, we aim to apply our framework to a larger-scale network and further investigate the application of multi-agent RL when facing multiple O/D pairs in the network.



\addtolength{\textheight}{-10cm}   








\bibliographystyle{ieeetr}
\bibliography{negar_bib.bib}

\end{document}